\documentclass[prd,twocolumn,showpacs,preprintnumbers]{revtex4}
\pdfoutput=1

\usepackage{amsfonts,amsmath,amssymb}
\usepackage{graphicx}
\usepackage{color}

\graphicspath{{./figs/}}

\begin{document}

\newcommand{\red}{\color{red}}
\newcommand{\blue}{\color{blue}}

\newcommand{\kp}{\kappa}
\newcommand{\Om}{\Omega}
\newcommand{\de}{\Delta}
\newcommand{\eps}{\epsilon}
\newcommand{\g}{\gamma}

\newcommand{\be}{\begin{equation}}
\newcommand{\ee}{\end{equation}}

\newcommand{\dec}{\mathrm{dec}}
\newcommand{\gsf}{g_{\ast s}^{\mathrm{f.o.}}}
\newcommand{\gd}{g_{\mathrm{D}}}
\newcommand{\gds}{g_{s, \mathrm{D}}}
\newcommand{\nast}{N_{\ast}}
\newcommand{\gast}{g_{\ast}}
\newcommand{\gasts}{g_{\ast s}}
\newcommand{\std}{\mathrm{std}}
\newcommand{\eff}{\mathrm{eff}}
\newcommand{\nr}{\mathrm{NR}}
\newcommand{\e}{\mathrm{end}}
\newcommand{\h}{\mathcal{H}}
\newcommand{\Hyd}{\mathrm{H}}
\newcommand{\He}{\mathrm{He}}
\newcommand{\eV}{\mathrm{eV}}
\newcommand{\keV}{\mathrm{keV}}
\newcommand{\GeV}{\mathrm{GeV}}
\newcommand{\lya}{Ly$\alpha \ $}
\newcommand{\iMpc}{\mbox{ Mpc$^{-1}$}}
\newcommand{\cms}{cm$^3$s$^{-1}$}
\newcommand{\sv}{\langle \sigma_{\mathrm{A}} v \rangle}
\newcommand{\ud}{\mbox{d}}
\newcommand{\fesc}{f_{\mathrm{esc}}}
\newcommand{\fa}{f_{\alpha}}
\newcommand{\fx}{f_{X}}
\newcommand{\Nion}{N_{\mathrm{ion}}}
\newcommand{\sigv}{\langle \sigma_A v \rangle }
\newcommand{\ic}{\mathrm{IC}}
\newcommand{\prompt}{\mathrm{prompt}}
\newcommand{\Mmin}{M_{\mathrm{min}}}
\newcommand{\nA}{n_{\mathrm{A}}}
\newcommand{\nH}{n_{\mathrm{H}}}
\newcommand{\nHe}{n_{\mathrm{He}}}
\newcommand{\Tgamma}{T_{\rm D}}

\newcommand{\apjl}{Astrophys. J. Lett.}
\newcommand{\apjs}{Astrophys. J. Suppl. Ser.}
\newcommand{\aap}{Astron. \& Astrophys.}
\newcommand{\aj}{Astron. J.}
\newcommand{\araa}{Ann. Rev. Astron. Astrophys. } 
\newcommand{\mnras}{Mon. Not. R. Astron. Soc.}
\newcommand{\physrep}{Phys. Rept.}
\newcommand{\jcap}{JCAP}

\newcommand{\sigmathermal}{\langle \sigma v\rangle}
\title{ Probing interactions within the dark matter sector via extra radiation contributions}
\author{Urbano \surname{Fran\c{c}a}}%
\author{Roberto A. \surname{Lineros}}%
\author{Joaquim \surname{Palacio}}%
\author{Sergio \surname{Pastor}}%

\affiliation{Instituto de F\'{\i}sica Corpuscular  (CSIC-Universitat de Val\`{e}ncia), 
Apdo.\ 22085, 46071 Valencia, Spain}
%


\begin{abstract}
The nature of dark matter is one of the most thrilling riddles for both 
cosmology and particle physics nowadays. 
While in the typical models the dark sector is composed only 
by weakly interacting massive particles, an arguably more natural scenario would include a whole
set of gauge interactions which are invisible for the standard model but that are in contact with 
the dark matter. We present a method to constrain the number of massless gauge bosons and other 
relativistic particles that might be present in the dark sector
using current and future cosmic microwave background data, 
and provide upper bounds on the size of the dark sector. 
We use the fact that the dark matter abundance depends on the strength of the interactions with both sectors,
which allows one to relate the freeze-out temperature of the dark matter with the temperature of {this cosmic background of dark gauge bosons}. This relation can then be used to calculate how sizable is the 
impact of the relativistic dark sector in the number of degrees of freedom of the early Universe,
providing an interesting and testable connection between cosmological 
data and direct/indirect detection experiments. 
The recent Planck data, in combination with other cosmic microwave background experiments and baryonic acoustic oscillations data, constrains the number of relativistic dark gauge bosons, when the freeze-out temperature of the dark matter is larger than the top mass, to be $N \lesssim 14$ for the simplest scenarios, while those limits are slightly relaxed for the combination with the Hubble constant measurements to  $N \lesssim 20$. Future releases of Planck data are expected to reduce the uncertainty by approximately a factor 3, what will reduce significantly the parameter space of allowed models.
\end{abstract}

\pacs{98.80.-k, 98.70.Vc, 95.35.+d, 98.80.Cq}
\preprint{IFIC/13-14}

\maketitle

\section{Introduction}
The standard model (SM) of particle physics has been extremely successful in describing accelerator and other 
terrestrial experiments over the last decades.
The recent discovery of a Higgs-like particle at the Large Hadron Collider~\cite{Higgs1, Higgs2}, for instance, seems to piece together a model that describes the fundamental interactions of nature using gauge theories. 
Nonetheless, cosmological data indicates that the matter content described by the SM are responsible for only around 4\% of the current energy density of the Universe~\cite{wmap9}, the remaining being in the form of the still unknown dark energy that drives the acceleration of the cosmos~\cite{darkenergy}, and dark matter (DM) which holds together galaxies and larger structures and corresponds to about 26\% of the energy budget of the Universe. 
Dark matter is assumed to be composed by particles, and typical candidates are in the form of weakly-interacting massive particles (WIMPs) that arise naturally in supersymmetric models~\cite{Jungman96}, extra--dimensions~\cite{Hooper:2007Ph,Cheng:2002PhR}, and extended Higgs sectors~\cite{Lopez:2007JC,Hirsch:2010ru}.
In analogy to what we observe in the visible sector of the Universe, the dark sector (DS) could be composed by more species, besides the DM candidate, including extra dark particles and interactions,
as for instance the case of models with hidden SU($N$) gauge symmetries.
This scenario would certainly lead to a richer phenomenology, if not in accelerators, at least in terms of the early and late Universe.
Those species could, for instance, freeze-out/decouple at different temperatures and leave an imprint in the early Universe, and/or could potentially explain the DM abundance in a different fashion than the so-called ``WIMP-miracle''~\cite{Bertone04,Feng10} where the DM abundance comes from the SM gauge and/or Higgs bosons exchange between DM and SM particles.
Other alternatives include the so-called mirror DM models~\cite{Foot:2004b,Foot:2004} where the interaction between DM and SM particles happens through the kinetic mixing between SM photons and dark photons, the latter coming from a hidden U(1) gauge symmetry interacting only with DM.
The strength of the mixing controls then the DM abundance, indirect, and direct detection signals.
Moreover, this mixing allows dark photons to slightly interact with SM charged particles, e.g. with electrons and positrons, modifying the evolution of the Universe~\cite{Ciarcelluti:2010}.
Nevertheless, such kinetic mixing is not present when the DS is composed by unbroken non-abelian gauge symmetries.
In this article we consider the implications that interacting WIMP--DS scenarios would have for the early Universe cosmology. 
We discuss how current cosmic microwave background (CMB) observations 
{from the Planck satellite~\cite{planck}, in combination with the Atacama Cosmology Telescope (ACT)~\cite{Sievers13}, the South Pole Telescope (SPT)~\cite{Hou12}, WMAP data~\cite{Calabrese13}, baryonic acoustic oscillations (BAO)~\cite{Percival10,Blake11,Padmanabhan12,Anderson13}, as well as with astrophysical measurements of the Hubble constant~\cite{Riess11,Freedman12}, can place limits on those new light particles within the dark sector. }
As we will show, if those ``dark radiation particles'' responsible for the DS interaction were in thermal equilibrium with the visible Universe via DM particles, they could potentially 
leave an imprint in the number of degrees of freedom (d.o.f.) of the early Universe that could be observed and/or constrained in the near future. 

While similar scenarios have been discussed before in the literature (see for instance~\cite{Feng08, Ackerman09, Blennow:2012, Fan:2013} and references therein), our discussion generalizes the calculations for potentially more realistic situations in which the interactions in the DS are mediated by more than one species that can have different freeze-out temperatures. 
Moreover, our results do not exclude other cosmological effects similar to the case of neutrinos in the visible sector, as the currently cosmological model requires at least three neutrinos with a temperature slightly lower when compared to photons, \mbox{$T_{\nu} \simeq (4/11)^{1/3} \, T_{\gamma}$}~\cite{Lesgourgues13}.

If those interactions take place within the dark sector only, cosmological observations like the ones discussed here will potentially be one of the few windows available to understand those new fundamental sectors.\\

\section{Early Universe Cosmology}
The evolution of the Universe during the radiation dominated epoch is described by the standard Friedmann equation~\cite{Kolb90,Lesgourgues13},
\begin{equation}
	H^2 = \frac{8\pi G}{3} \, \rho_r , 
\end{equation}
where $H$ is the Hubble rate and $G$ the Newton constant. The energy density in radiation after the annihilation of electron-positron pairs is given by
\begin{equation}
	\rho_r = \rho_{\gamma} \left[ 1+ \frac{7}{8} \left( \frac{11}{4} \right)^{4/3} \times N_{\eff} \right] \, .
\end{equation}
Here the {\it effective number of neutrinos} is given by \mbox{$N_{\eff} = N_{\eff}^{\std} + \Delta N_{\eff}$}, where $N_{\eff}^{\std} = 3.046$ is the standard contribution from the three SM neutrinos~\cite{Mangano05, Lesgourgues06, Signe13}, and $\Delta N_{\eff}$ is the contribution coming from extra light species present in the early Universe, known in the literature as {\it dark radiation}, like e.g.~sterile neutrinos \cite{Giusarma11, Hamann11,Ho12,Ho13,Higaki13}, or from other physical processes, like lepton asymmetries~\cite{Mangano10, Castorina12}, reheating of neutrinos~\cite{Boehm12}, gravitational waves~\cite{Sendra12}, and other effects~\cite{Steigman13}.
Here, for simplicity of the analysis, we will assume that those {other possible} contributions are negligible, and any extra contribution for $N_\eff$ comes from the extra d.o.f.\ stored in the dark sector.\\

\section{Degrees of freedom of the dark sector}
To account for the d.o.f.\ in the dark sector and calculate the contribution to the effective number of neutrinos, we
will adopt the standard tools for calculating the entropy and gravitational impact of the new particles in the early Universe~\cite{Kolb90}. 
We assume that those new particles are either massless or very light, and therefore relativistic, when their interaction rate $\Gamma_D$ in the dark sector became smaller than the Hubble rate $H^{-1}$, effectively freezing-out their distribution as bosons or fermions. 
In our analysis the dark sector can be composed by $N$ species of dark gauge bosons (DBs), i.e. massless spin-1 particles, that mediate the interactions between dark matter particles 
and $M$ species of relativistic dark fermions (DFs) with different temperatures $T_i = x_i \, \Tgamma$ and masses $m_i \ll x_i T_D$. 
The latter is inspired by the visible sector, in which neutrinos and photons have different temperatures.
$\Tgamma$ is the temperature of the background of {dark gauge bosons}, which is lower
than the temperature of visible photons as the latter get entropy contributions
from particle-antiparticle annihilations.
To calculate this extra contribution to $N_\eff$, 
{ one can use the entropy conservation to estimate the ratio of temperatures in both sectors. Since the entropy is given by $S= g_{\ast s} a^3 T^3$, being $a$ the scale factor, and under the approximation that all the decouplings occur instantaneously, }
the ratio of temperatures can be written as
\begin{equation} \label{eq:ratioT}
	\frac{T_i}{T_{\gamma}} = x_i \frac{\Tgamma} {T_{\gamma}} \approx x_i \left(\frac{g_{\ast s 0}}{\gsf}\right)^{1/3} \ ,
\end{equation}
where $\gsf$ is the number of d.o.f. of the Universe when the interactions that maintained the dark (relativistic) and visible sectors in thermal equilibrium ceased, and $g_{\ast s 0}$ corresponds to its current value. 
For thermal DM candidates, $\Tgamma$ can be related to the DM freeze-out temperature $T_{\rm DM}^{\rm f.o.}$ and to the dark matter mass  $M_{\rm DM}$. 
In the case of WIMPs, for instance, we have that
\begin{equation}
	T_{\rm DM}^{\rm f.o.} \approx \frac{1}{20} \, M_{\rm DM} \, ,
\end{equation}
is a fraction of the DM mass~\cite{Beacom12}, and therefore one could go a step further and relate $\Tgamma$ to $M_{\rm DM}$, a feature that holds for any model in which the DM is the only link between the dark and visible sectors. 
In addition to that, the observed DM abundance~\cite{planck}, 
\begin{equation}
	\Omega_{\rm DM} h^2 = 0.1196 \pm 0.0031 \, ,
\end{equation}
requires a precise value (to the percent level) of the thermally averaged scattering cross-section \mbox{$\sigmathermal \simeq 3 \times 10^{-26} \, {\rm cm}^3 {\rm s}^{-1}$}, what fixes the relation among DM and its interactions with the visible and dark sectors,
\begin{equation}
	\label{eq:xsec}
	\sigmathermal = \sigmathermal_{\rm SM} + \sigmathermal_{\rm DS} \simeq 3 \times 10^{-26} \, {\rm cm}^3 {\rm s}^{-1}\, ,
\end{equation}
and this, in turn, can be translated into a relation between the dark sector decoupling and dark matter freeze-out temperatures~\cite{Kolb90},
\begin{equation}
	\label{eq:temps}
\frac{\Tgamma}{T_{\rm DM}^{\rm f.o.}} \simeq \frac{ 
		18 + \log{\left( \displaystyle \frac{\sigmathermal}{3 \times 10^{-27} \, {\rm cm}^3 {\rm s}^{-1}} \, \frac{M_{\rm DM}} {\rm GeV}  \right)}}
		{ 18 + \log{\left( \displaystyle \frac{\sigmathermal_{\rm DS}}{3 \times 10^{-27} \, {\rm cm}^3 {\rm s}^{-1}} \, \frac{M_{\rm DM}}{\rm GeV}  \right)}} \, .
\end{equation}
The case when dark radiation particles cannot communicate directly with the visible sector allows DM to help dark radiation particles to be in thermal equilibrium with photons up to the time of DM freeze-out. 
This condition connects the temperature of DM freeze-out with $\Tgamma$ and roughly implies a lower bound on $\Tgamma$ set by $T_{\rm DM}^{\rm f.o.}$ when the DM annihilation cross-section is dominated by DS interactions, as can be seen from Eq.~(\ref{eq:temps}). {This case, $\sigmathermal_{\rm DS} \simeq \sigmathermal$, would reduce the posibility of detection of WIMPs in direct/indirect searches. On the other hand, a scenario with WIMP--detection prospects would be: $\sigmathermal_{\rm DS} \sim 10^{-3} \times \sigmathermal$, which increases the decoupling temperature to $\Tgamma \sim 1.5 \times T_{\rm DM}^{\rm f.o.}$ due to the logarithmic dependence shown in Eq.~(\ref{eq:temps}).}
This scenario can be well motivated in the context of particle physics models where there exist a hidden gauge symmetry like SU($N$) symmetries.
These hidden symmetries can reduce the interaction strength between DM and the SM and, as consequence, the indirect and direct detection signal. 
This feature is important to avoid exclusion regions set by direct detection experiments like XENON100~\cite{Xenon100:2011}. 
In a similar fashion, asymmetric DM models (that aim to explain why DM and baryon abundances are similar) can affect the evolution of the early Universe when interactions with the dark radiation particles are included~\cite{Blennow:2012}. 
However, contrary to the WIMP case, for those scenarios the mechanism of DM production and freeze-out is usually model dependent, and therefore a general study is more challenging and probably not feasible.
In the scenarios discussed here, the main prediction is the connection of the freeze-out temperature with the extra d.o.f. that will impact $N_{\eff}$ as, for instance in eq.\ (\ref{eq:temps}), that we will discuss in detail elsewhere~\cite{longpaper}. 
Using the definition of entropy d.o.f. and eq.(\ref{eq:ratioT}) one can write $g_{\ast s 0}$ as 
\be
g_{\ast s 0} \approx 3.91 + \gds \left( \frac{g_{\ast s 0}}{\gsf} \right) \ ,
\ee
where 
\begin{equation}
	\label{eq:gds}
	\gds  \equiv  \sum_{i}^N g_i x_i^3 + \frac{7}{8} \sum_{j}^M g_j x_j^3 \ , 
\end{equation}
and therefore,
\begin{equation}
	g_{\ast s 0} \approx 3.91 \, \frac{\gsf}{(\gsf - \gds)} \ .
\end{equation} 
The increment in $N_\eff$ due to the new particles in the dark sector can then be written as
\begin{equation}
	\label{eq:DeltaGeneral}
	\Delta N_{\eff} \approx 2.201 \ \gd \left(\frac{3.91}{\gsf - \gds} \right)^{4/3} \ ,
\end{equation}
where 
\begin{equation}
	\label{eq:gsFGeneral} 
	\gd \equiv  \sum_{i}^N g_i x_i^4 + \frac{7}{8} \sum_{j}^M   g_j x_j^4 \ .
\end{equation}
Notice that the factor $3.91$ comes from the number of entropy d.o.f. after the $e^{+} e^{-}$ annihilations.\\

\section{Discussion}
We will discuss here two particular toy models that simplify the equations above with the goal to learn what kind of constraints can currently be placed in those scenarios using CMB data (in combination with other cosmological observations).
First, we will describe a scenario in which all $N$ massless ``dark gauge bosons'' have the same temperature, and it is associated with the dark matter mass via the freeze-out temperature of DM. For simplicity, we assume that all DBs belong to the same family of the dark particle background i.e. $x_{{\rm DB},i} = 1$.
After that, we will generalize this scenario by allowing for $M$ massless ``dark fermions'' that have a slightly lower temperature than the dark bosonic sector, a situation that mimics the photon and neutrino backgrounds of the visible sector.

In the case of $N$ dark gauge bosons, eqs. (\ref{eq:gds}), (\ref{eq:DeltaGeneral}), and (\ref{eq:gsFGeneral}) connect the three main quantities as
\begin{equation}
	\label{eq:NeffB}
	\Delta N_{\eff} \approx 4.402 \, N \, \left( \frac{3.91}{\gsf - 2 N} \right)^{4/3} .
\end{equation}

\begin{figure}[tb]
	\centering
	\includegraphics[width=\columnwidth]{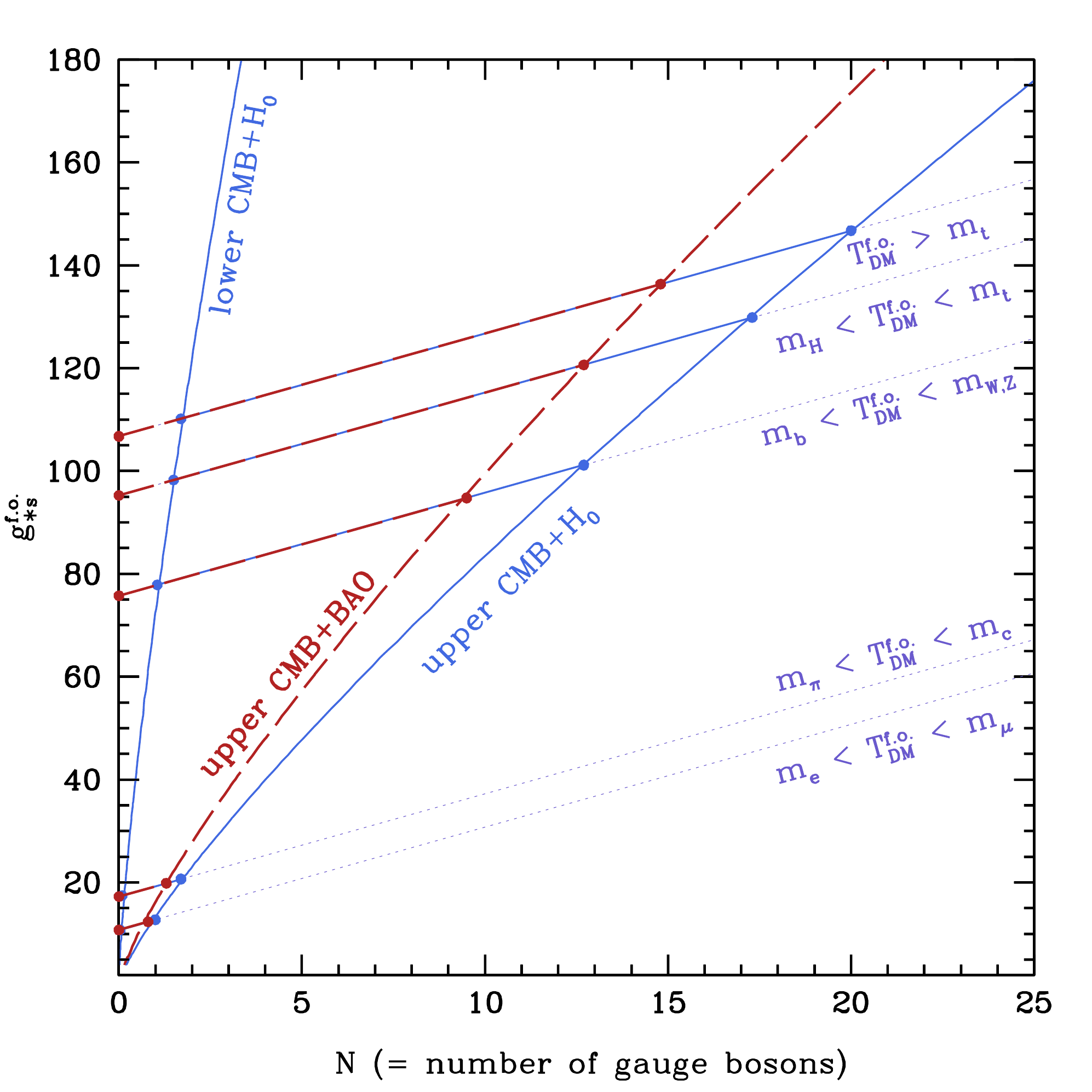}
	\caption{\label{fig:neffb} 
		Number of d.o.f. $\gsf$ versus number of dark gauge bosons.
		The red dashed line corresponds to the current 95\% C.L.~upper value on $N_{\eff}$ using 
CMB+BAO~\cite{planck}, while the upper/lower blue solid lines shows the combination CMB+$H_0$ (see text for details). 
		Diagonal dotted lines represent $\gsf$ levels associated to $\Tgamma$ for the SM particle content. Notice that for a given range of $T_{\rm DM}^{\rm f.o.}$, models exist only along the corresponding line.}
\end{figure}

In this case, $\gsf$ and $N$ have a non--null impact on $N_{\eff}$ except in trivial cases like $N = 0$ or extremely large $T_i$ i.e. $\gsf \rightarrow \infty$. 
{The implications for $N_\eff$ are presented in Fig.~\ref{fig:neffb} where the lines show two of the 95\% C.L.  
constraints quoted in Ref.~\cite{planck}: from CMB (where CMB corresponds to the combination of data from Planck, WMAP--Polarization, SPT, and ACT) on $N_{\eff}$~\cite{planck}, in combination with {\it (i)} BAO~\cite{Percival10,Blake11,Padmanabhan12,Anderson13}, 
\be
N_{\eff} = 3.30^{+0.54}_{-0.51} \ \ \ (95\% \, {\rm C.L.}; \mathrm{CMB+BAO}),
\ee
and in combination with
{\it (ii)} the measurement of the Hubble constant $H_0$~\cite{Riess11,Freedman12}. 
\be
N_{\eff} = 3.62^{+0.50}_{-0.48} \ \ \ (95\% \, {\rm C.L.}; \mathrm{CMB+}H_0).
\ee
Notice that the latter
excludes at 95\% C.L. the standard value of $N_{\eff}$, requiring the presence of extra radiation, such as the dark sector.}
The diagonal (light dotted) lines represent the number of {\it total} degrees of freedom (visible and in the dark sector) at the freeze-out epoch, $\gsf = \gsf |_{{\mathrm SM}}+\gsf |_{{\mathrm DS}}$.

\begin{figure}[tb]
	\centering
	\includegraphics[width=\columnwidth]{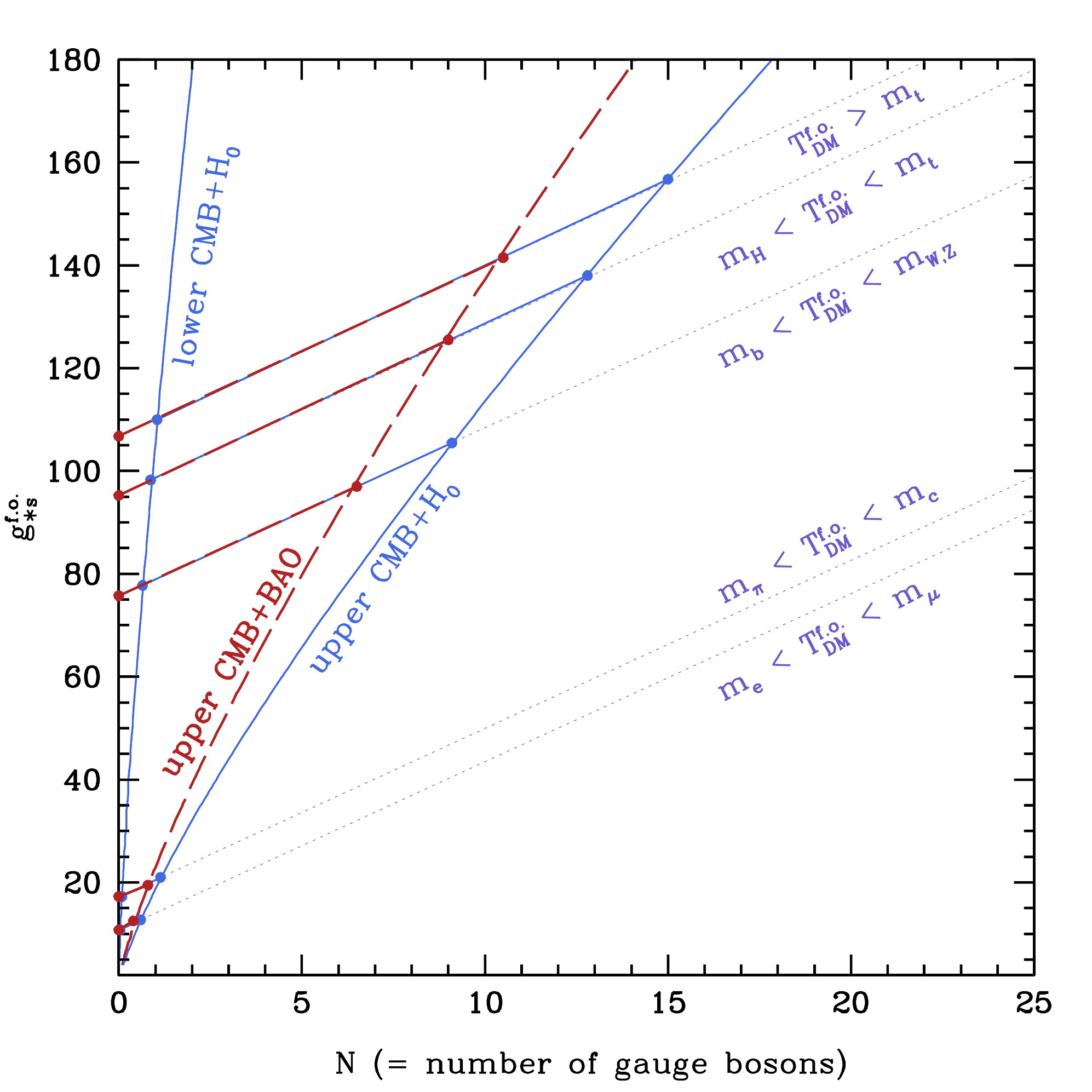}
	\caption{\label{fig:neffbf}
		Number of d.o.f.\ $\gsf$ versus number $N $ of dark gauge bosons for the case with additional $3 N$ dark fermions. 
		Line-styles are equivalent to the ones shown in Fig.~\ref{fig:neffb}.}
\end{figure}

Depending on the decoupling temperature of DBs, we can infer the maximum size of the dark sector allowed by the uncertainties in $N_\eff$. 
For instance, all models for which $T_{\rm DM}^{\rm f.o.}$ is larger than the top mass ``live'' in the line indicated by $T_{\rm DM}^{\rm f.o.} > m_t$, and therefore current constraints (dashed line) on $N_{\eff}$ present an upper limit on
the number of dark gauge bosons of approximately 14 (CMB+BAO datasets).
For temperatures between the bottom--quark and the $W, Z$ bosons masses, the bound is reduced to $\approx 10$ DBs.  
Smaller values of $T_{\rm DM}^{\rm f.o.}$ shrink the dark sector down to few or no DBs, although we should keep in mind that sub-GeV temperatures are affected by the QCD phase transition. 

The addition of {$M = 3 N$} dark fermions affects the value of $N_\eff$ by
\begin{equation}
	\label{eq:NeffBF}
	\Delta N_{\eff} \approx 7.369 \, N \, \left( \frac{3.91}{\gsf - 3.91 \, N} \right)^{4/3} \, ,
\end{equation}
which is similar in form to the case composed only by DBs. 
In Fig.~\ref{fig:neffbf}, we present the implications for this setup. 
As one would expect, the presence of DFs leads to a reduction on the maximum number of DBs because they also contribute to the radiation density. 
For the freeze-out temperatures discussed in the previous case, we obtain an upper bound of approximately 10 and 9 dark gauge bosons for $T_{\rm DM}^{\rm f.o.} > m_{t}$ and $m_{b} < T_{\rm DM}^{\rm f.o.} < m_{W,Z}$, respectively for current data. The expected constraints that one can place on this scenario are expect to improve in the near future as CMB experiments continue to take data, and we expect that future data releases might rule out whole classes of models.

We should emphasize that the dark sector also contains one or more stable (or quasi-stable) massive states that play the role of DM. While those particles have a negligible contribution to our calculations in terms of radiation, they { have governed} the evolution of the Universe for a long time and are responsible for the formation of structures. 
For simplicity, we consider the case of WIMPs, which is a widely studied DM candidate with well-behaved thermal history.
If the DS interaction rate is similar to the SM interaction rate, Eq.~(\ref{eq:xsec}) implies a relation between $T_{\rm DM}^{\rm f.o.}$ and $M_{\rm DM}$. In particular, WIMP masses in the range $100$ GeV -- $1800$ GeV constrain the maximum number of DBs to $\approx 14 \ (20)$  at present, depending on the datasets.
While this analysis was done under the assumption that the dark and visible sectors shared the same temperature at some point, it should 
also work for models that were never in thermal equilibrium with SM particles but for which the couplings of the inflaton with both the dark and visible sector are known: in this case one can show that $\Tgamma$ will be related to the reheating temperature and to the couplings of the inflaton with each species~\cite{Feng08}, and they can be translated into different $x_i$. 
Yet another alternative is to think of the temperature ratios $x_i$ as free parameters instead of being given by the particle physics model, and use future precise data to constrain them, although in this case one would most probably need extra datasets to disentangle the correlations between $x_i$ and the number of particles in the dark sector.\\

\section{Conclusions}
The dark sector is generally assumed to be composed only by 
a WIMP dark matter, but a feasible (and arguably more natural) model  would
include dark gauge bosons and other particles that might be stable. 
In this article we discussed what is potentially 
one of the very few windows to place constraints on 
the number and interactions of those particles, and applied our
general results to a couple of toy models to understand to what 
level current and future cosmological data can constrain those scenarios. 

In particular, we showed that for models in which  the DM is weakly interacting, one
can relate its freeze-out temperature to the temperature of the rest of the
dark sector, and therefore amend the standard analysis of d.o.f.\ in the early 
Universe to take into account the contribution of this dark sector to the
effective number of neutrinos. Moreover, in the context of the
standard cosmological model, our method can also be easily generalized for cases 
in which the dark sector was not in thermal  equilibrium with SM particles, since
both sectors are supposed to have a common origin after the inflationary phase.

We calculated this extra contribution
to $N_{\eff}$ in an almost model independent way, and particularized those results
for some simple models to compare them with current and future data. In particular, we
obtained the constraints on the number of interactions within the dark sector:
 while current limits on the number of
dark gauge bosons are somewhat weak ($\lesssim 14 \ (20)$), 
{future CMB data will  potentially improve those
bounds by a factor of approximately 3 (assuming they can improve sensitivity 
to $\sigma[N_{\eff}] \approx 0.2$ at 95\% C.L.), }
ruling out
whole classes of models. {It will be interesting to look
at other WIMP DM models that can be accommodated
within this scenario and constrain their properties with current and 
forthcoming cosmological data.}

\section*{Acknowledgments}
We thank M.~Hirsch, M.~Malinsk\'{y}, and M.~Taoso for useful discussions. This work was supported by the Spanish MINECO under grants FPA2011-22975 and MULTIDARK CSD2009-00064 (Consolider-Ingenio 2010 Programme), by Prometeo/2009/091 (Generalitat Valenciana), and by the EU ITN UNILHC PITN-GA-2009-237920.

\end{document}